\documentclass{ifacconf}

\usepackage{natbib}        
\usepackage{amsmath}
\usepackage{amssymb}
\usepackage{mathtools}

\usepackage{graphicx}
\usepackage{color}
\usepackage[usenames]{xcolor}

\usepackage[hyphens]{url}	

\usepackage{siunitx}
\usepackage{tikz}
\usetikzlibrary{
    angles,
    arrows,
    arrows.meta,
    automata,
    calc,
    fit,
    patterns,
    positioning,
    quotes,
    shapes,
}
\usepackage{pgfplots}
\pgfplotsset{compat=newest}
\tikzstyle{block} = [draw, thick, node distance=1cm, minimum width=1cm, inner sep=6pt]

\usepackage{wrapfig}
\usepackage{nicefrac}
\usepackage{subfig}

\newcommand{\lineannShort}[4][0.5]{%
	\begin{scope}[rotate=#2, inner sep=2pt]
		\coordinate (a) at (0,0);	
		\coordinate (b) at (#3,0);	
		\draw[dashed] (b) -- +(0,#1);
		\draw[|<->|] (a) -- node[rotate = #2, above] {#4} (b);
	\end{scope}
}

\renewenvironment{thebibliography}[1]{%
    \begin{oldthebibliography}{#1}%
    \setlength{\parskip}{0ex}%
    \setlength{\itemsep}{0ex}%
}%
{%
    \end{oldthebibliography}%
}
\newtheorem{procedure}{Procedure}
\begin{document}
\setlength{\abovedisplayshortskip}{0.5ex plus1ex minus1ex}
\setlength{\abovedisplayskip}{0.5ex plus1ex minus1ex}
\setlength{\belowdisplayshortskip}{0.9ex plus1ex minus1ex}
\setlength{\belowdisplayskip}{0.9ex plus1ex minus1ex}
\begin{frontmatter}

\title{Collision Avoidance Safety Filter for an Autonomous E-Scooter using Ultrasonic Sensors\thanksref{footnoteinfo}} 

\thanks[footnoteinfo]{%
F.\ Allgöwer is thankful that this work was funded by the Ministry of Science, Research and the Arts of the State of Baden-Württemberg (MWK) in the context of the ``MobiLab'' Project.
R.\ Strässer, D.\ Meister, M.\ Seidel thank the Graduate Academy of the SC SimTech for its support.
F.\ Brändle thanks the International Max Planck Research School for Intelligent Systems (IMPRS-IS) for its support.
\\© 2024 the authors. This work has been accepted to IFAC for publication under a Creative Commons Licence CC-BY-NC-ND.\\[-2.75\baselineskip]
}

\author[First]{Robin Strässer}
\author[First]{Marc Seidel}
\author[First]{Felix Brändle} 
\author[First]{David Meister}
\author[Second]{Raffaele Soloperto}
\author[Third]{David Hambach Ferrer}
\author[First]{Frank Allgöwer}

\address[First]{University of Stuttgart, Institute for Systems Theory and Automatic Control, 70550 Stuttgart, Germany
		(e-mail: e-scooter@ist.uni-stuttgart.de)}
\address[Second]{ETH Zürich, Automatic Control Laboratory, 8092 Zürich, Switzerland}
\address[Third]{B.Sc. student at the University of Stuttgart, 70550 Stuttgart, Germany}

\begin{abstract}                
    In this paper, we propose a collision avoidance safety filter for autonomous electric scooters to enable safe operation of such vehicles in pedestrian areas.
    In particular, we employ multiple low-cost ultrasonic sensors to detect a wide range of possible obstacles in front of the e-scooter.
    Based on possibly faulty distance measurements, we design a filter to mitigate measurement noise and missing values as well as a gain-scheduled controller to limit the velocity commanded to the e-scooter when required due to imminent collisions.
    The proposed controller structure is able to prevent collisions with unknown obstacles by deploying a reduced safe velocity ensuring a sufficiently large safety distance.
    The collision avoidance approach is designed such that it may be easily deployed in similar applications of general micromobility vehicles.
    The effectiveness of our proposed safety filter is demonstrated in real-world experiments.
\end{abstract}


\end{frontmatter}
 
\section{Introduction}
\vspace*{-0.75\baselineskip}

Sharing systems for electric scooters (e-scooters) are well established in many cities around the globe~\citep{Goessling2020}.
E-scooters are a popular choice for urban mobility, particularly for short-distance commutes~\citep{Degele2018}. 
While on-demand availability and distribution in dockless sharing systems are part of their appeal, these are also the cause of various practical challenges~\citep{Hollingsworth2019,Tuncer2020,Goessling2020}.
E-scooter related issues include cluttered sidewalks due to inconveniently parked or dropped e-scooters, as well as the need for additional vehicles and personnel to collect and charge the e-scooters or relocate them.
To achieve high on-demand availability, suppliers often rely on non-sustainable and staff-intensive relocation of e-scooters using large vehicles.
In addition, a large number of e-scooters are typically deployed to increase availability.
These challenges result in high operational costs - economically~\citep{Heineke2019,Rose2020}, ecologically~\citep{Cazzola2020,Krauss2022}, and socially~\citep{Farley2020,Gossling2020,Gioldasis2021,Mehdizadeh2023}.
\vspace*{-0.25\baselineskip}

Given these observations,~\cite{Wenzelburger2020} propose the development of an autonomous e-scooter that can resolve or alleviate many of the described concerns about shared e-scooters.
In particular, with the term \emph{autonomous e-scooter}, we refer to a standard two-wheeled e-scooter, augmented with additional features that enable the system to self-stabilize its upright position and autonomously navigate in a specified environment while avoiding unknown obstacles.
By endowing e-scooters with the capability to move autonomously while not in use, cost-intensive manual redistribution or charging become obsolete.
In addition, a ride with such an e-scooter is allowed to end anywhere within the region of interest, since this modified free-floating sharing system leverages the advantages of station-based sharing systems by autonomously parking an e-scooter after its ride and, thus, preventing cluttered public spaces.
Moreover, their autonomy allows for optimal self-distribution according to the current or forecasted demand which improves availability and, thereby, allows to reduce the number of e-scooters necessary to cover an area of operation.
\vspace*{-0.25\baselineskip}

\emph{Related work:} 
As the availability and popularity of micromobility vehicles is rapidly increasing, also the proper interaction of such vehicles with their environment is of high interest, see, e.g.,~\cite{Li2023}.
Further, autonomous operation in micromobility needs to be carefully designed due to the close interaction with pedestrians and other parts of the urban environment~\citep{Christensen2021}.
Autonomous cars typically rely on a variety of expensive sensors~\citep{Koon2023}, whereas micromobility vehicles are often designed with a lower budget and therefore have only a few low-cost sensors installed.
Although ultrasonic sensors are known to be of low cost in installation and post-processing compared to, e.g., LiDAR sensors or cameras, they are rarely used in micromobility for obstacle detection or collision avoidance~\citep{Manikandan2022,Guglielmi2023}. 
However, ultrasonic sensors already proved useful, e.g., for the obstacle avoidance of robots~\citep{Borenstein1988,Yasin2020} or lateral collision avoidance of cars~\citep{Song2004}. 
\vspace*{-0.25\baselineskip}

\begin{figure}[bt]
    \vskip -0.025 in
    \centering
	\includegraphics[width = 0.45\linewidth]{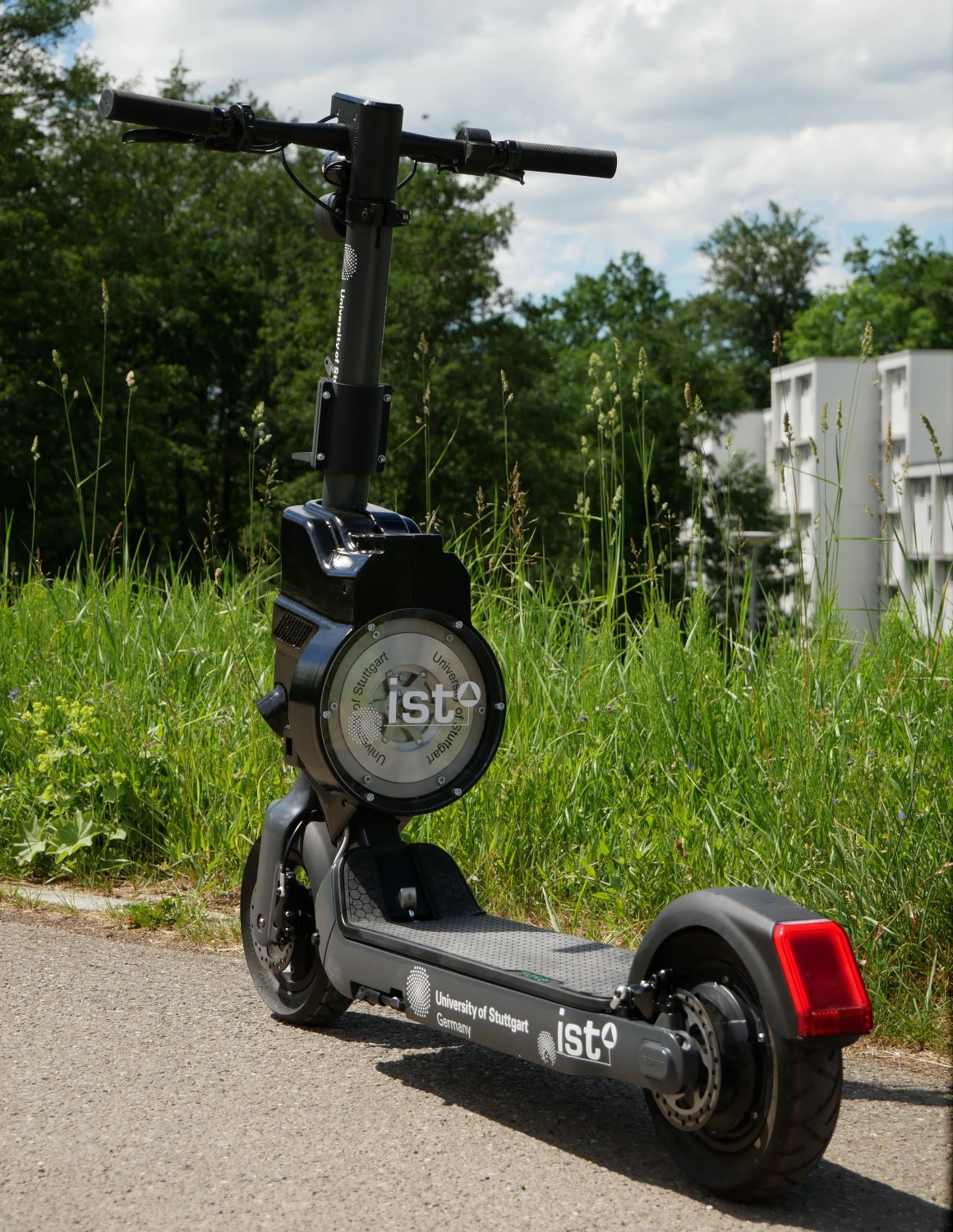}
	\vspace*{-0.75\baselineskip}
	\caption{The developed autonomous e-scooter prototype.}
	\vspace*{-0.25\baselineskip}
	\label{fig:picture_escooter}
\end{figure}

\emph{Contribution:} 
In this paper, we focus on the development of a collision avoidance safety filter for the autonomous e-scooter shown in Fig.~\ref{fig:picture_escooter} using low-cost ultrasonic sensors.
The main contribution of our proposed solution is the integration of a feedback design for safety stops.
More precisely, we propose a gain-scheduled controller that limits the velocity applied to the autonomous e-scooter based on filtered distance measurements of the sensors.
Here, it is crucial to design a suitable filter for the sensors to account for faulty or missing measurements.
Since the e-scooter is not only driving straight but also in curves, we need to ensure that the controller is able to handle different scenarios. 
To this end, we incorporate three different ultrasonic sensors, where one faces straight and the other two are oriented to the sides of the e-scooter.
The designed feedback solution allows to weigh the measurements of all three sensors according to the currently applied steering angle to ensure that only obstacles which block the imminent path are considered in the control decision.
The proposed approach is designed to be easily transferrable to other autonomous vehicles, where the focus is the deployment in pedestrian areas with a low-cost design and, thus, low complexity requirement.
Further, we aim to provide an easy-to-follow implementation and design guide for general applications.
To this end, we include a description of the hardware setup as well as the respective controller design.
\vspace*{-1.35\baselineskip}

\emph{Outline:} 
In Section~\ref{sec:hardware}, we provide an overview of the hardware setup of the autonomous e-scooter, and describe the dynamics of the e-scooter.
Section~\ref{sec:controller} details the proposed collision avoidance safety filter.
The derived approach is demonstrated in Section~\ref{sec:experiments} by means of real-world experiments.
Finally, we conclude the paper in Section~\ref{sec:conclusion}.
\vspace*{-0.3\baselineskip}
\section{System Overview}\label{sec:hardware}
\vspace*{-0.8\baselineskip}

In this section, we provide an overview of the developed e-scooter prototype. 
We describe the system dynamics in Section~\ref{sec:system-dynamics}, and the hardware setup in Section~\ref{sec:hardware-setup}, where we elaborate specifically on the added components for the proposed collision avoidance safety filter.
\vspace*{-0.5\baselineskip}

\subsection{System dynamics}\label{sec:system-dynamics}
\vspace*{-0.65\baselineskip}

\begin{figure}[t]
    \vskip -0.075 in
    \subfloat[Ultrasonic sensors, steering motor,\\and reaction wheel.]{\label{IMG:HW:EScooter}
        \begin{minipage}{0.7\linewidth}
            \centering
            \hspace*{-0.1\linewidth}
            \input{img/IMG_ScooterLabeled.tex}
        \end{minipage}
    }
    \hspace*{-0.075\linewidth}
    \subfloat[3D-printed ultrasonic sensor mount.]{\label{IMG:HW:ultrasonic}
        \begin{minipage}{0.3\linewidth}
            \centering
            ~\\[1\baselineskip]
            \includegraphics[width=1.16\linewidth]{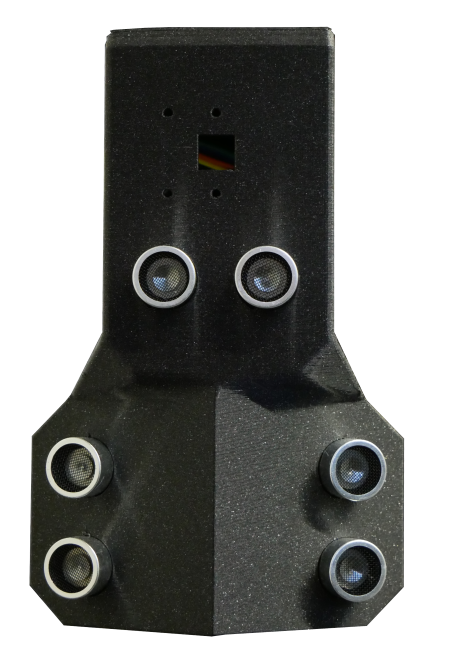}
        \end{minipage}
    }
    \vspace*{-0.6\baselineskip}
	\caption{The modified e-scooter with additional components.}
    \vspace*{-0.45\baselineskip}
\end{figure}

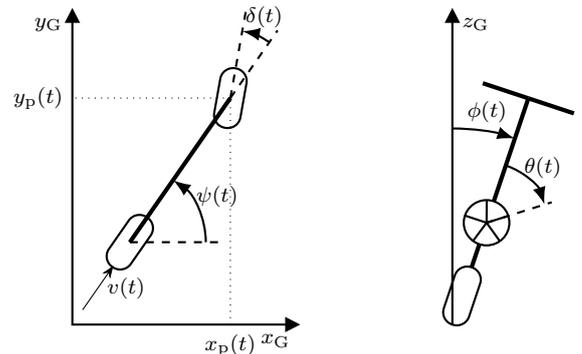
\begin{figure}[t]
    \centering
    \begin{minipage}{0.95\linewidth}
        \centering
        \usetikzlibrary{calc,angles,quotes,arrows.meta}

\tikzset{
pill/.style={minimum width=0.8cm,minimum height=3.5mm,rounded corners=1.7mm,draw}
}

\begin{tikzpicture}
\draw[thick,{Triangle[length=2mm]}-{Triangle[length=2mm]}] 
		(0,4.2)	coordinate (Y) node[below left]{\small$y_{\mathrm{G}}$} 
	--	(0,0)	coordinate (O)
	--	(3.0,0)	coordinate (X) node[below left] {\small$x_{\mathrm{G}}$}
;

\path
		(O) 
	--	(2.7,3.9) 
				coordinate[pos=0.05] (velHelper1) 
				coordinate[pos=0.20] (velHelper2) 
				coordinate[pos=0.28] (F1)	
				coordinate[pos=0.77] (F2)	
				coordinate(TR)				
;

\draw[-Stealth] (velHelper1) -- node[right]{\small$v(t)$}(velHelper2);

\draw[thick,dashed] (F2)--(TR);

\draw[ultra thick] 
		(F1)
	--	(F2)
;
\draw[thick] 
		(F1)
	--	(F2)
		node[pos=0,sloped,pill]{}
;
\draw[thick,dashed]
		(F2)
	-- ++ (80:1.2) coordinate(H)			
		node[pos=0,sloped,pill,solid]{}
	pic ["\small$\delta(t)$",draw,solid,-{Latex},angle radius=0.9cm,angle eccentricity=1.3] {angle = TR--F2--H}
;

\coordinate (F1R) at ($(F1) + (1.2,0)$);	
\draw[thick,dashed]
	pic ["\small$\psi(t)$",draw,solid,-{Latex},angle radius=1.cm,angle eccentricity=1.3] {angle = F1R--F1--F2}
;
\draw[thick, dashed] (F1) -- (F1R);

\draw[dotted] (F2 |- O) node[below]{\small$x_\mathrm{p}(t)$} -- (F2 |- F2);
\draw[dotted] (F2 |- F2) -- (O |- F2) node[left]{\small$y_\mathrm{p}(t)$};

\begin{scope}[shift={(5,0)}]
    \coordinate (Y) at (-1.5,0);
    \coordinate (O) at (0,0);
    \coordinate (Z) at (0,4.2);
    
    \tikzset{
    pill/.style={minimum width=2\wheelradius,minimum height=3.5mm,rounded corners=1.7mm,draw,fill=white}
    }
    
    \newlength{\wheelradius}
    \setlength{\wheelradius}{4mm}
    \newlength{\flywheelradius}
    \setlength{\flywheelradius}{3mm}
    
    \draw[ultra thick]
    		(O) 
    	--	(1.0,3.0)
    				coordinate	(HB)
    				coordinate[pos=0.45] (FW)	
    ;
    \draw[thick, fill=white] (FW) circle (3mm)
    ;
    \draw[ultra thick]
    		(HB) 
    	-- ($(HB)!0.2!90:(O)$) coordinate(HBR)	
    ;
    \draw[ultra thick]
    		(HB) 
    	-- ($(HB)!0.2!-90:(O)$) coordinate(HBL)	
    ;
    \draw[thick]
    		(FW) 
    	-- ($(FW)!\flywheelradius!-54:(HB)$) coordinate(FWT)	
    				coordinate[pos=3] (FWTE)				
    ;
    \draw[thick]
    		(FW) 
    	-- ($(FW)!\flywheelradius!-126:(HB)$)
    	;
    \draw[thick]
    		(FW) 
    	-- ($(FW)!\flywheelradius!-198:(HB)$)
    	;
    \draw[thick]
    		(FW) 
    	-- ($(FW)!\flywheelradius!-270:(HB)$)
    	;
    \draw[thick]
    		(FW) 
    	-- ($(FW)!\flywheelradius!-342:(HB)$)
    	;
    \draw[thick, dashed]
    		(FW)
    	--	(FWTE)
    	pic ["\small$\theta(t)$",draw,solid,{Latex}-,angle radius=0.8cm,angle eccentricity=1.3] {angle = FWTE--FW--HB}
    ;
    \draw[thick]
    		(O)
    	--	($(O)!\wheelradius!0:(HB)$)
    		node[pos=1,sloped,pill]{}			
    ;
    \draw[thick,-{Triangle[length=2mm]}]
    	(O)	
    	--	(Z) node[below right] {\small$z_{\mathrm{G}}$}
    ;
    \draw[thick,dashed]
    	pic ["\small$\phi(t)$",draw,solid,{Latex}-,angle radius=2.6cm,angle eccentricity=1.1] {angle = HB--O--Z}
    ;
\end{scope}

\end{tikzpicture}
    \end{minipage}
    \vspace*{-\baselineskip}
    \caption{Top view (left, single-track model) and rear view (right, inverted pendulum) on the e-scooter for illustration of the state vector $x(t)$, where $\dot{\theta}(t) = \omega(t)$.}
    \label{fig:states}
    \vspace*{-0.1\baselineskip}
\end{figure}

In the following, we describe the autonomous e-scooter to provide a better understanding of the studied prototype. 
A main feature of our prototype is that it balances itself on two wheels when moving autonomously, i.e., when not being used by a human.
As described in~\cite{Wenzelburger2020}, this is realized by a reaction wheel mounted between deck and stem in order to self-stabilize the e-scooter, see Fig.~\ref{IMG:HW:EScooter}.
From a dynamical systems perspective, the e-scooter is described by the state vector
$
    x(t)
    \coloneqq 
    \begin{bmatrix}
        x_{\mathrm{p}}(t) 
        & y_{\mathrm{p}}(t) 
        & \psi(t)
        & v(t)
        & \delta(t)
        & \phi(t)
        & \dot{\phi}(t)
        & \omega(t)
    \end{bmatrix}^\top
$,
where $x_{\mathrm{p}}(t)$, $y_{\mathrm{p}}(t)$, and $\psi(t)$ represent the position and orientation of the e-scooter, respectively, $v(t)$ is the (linear) velocity at the rear wheel, and $\delta(t)$ denotes the steering angle for some time $t\geq 0$.
In addition, we have three states corresponding to the roll dynamics of the e-scooter: the roll angle $\phi(t)$, its derivative $\dot{\phi}(t)$, and the motor velocity of the reaction wheel $\omega(t)$.
The physical meaning of the individual components of the state vector is illustrated in Fig.~\ref{fig:states}.
Further, the system is controlled via the input vector
$
    u(t) \coloneqq \begin{bmatrix} 
        v_{\mathrm{cmd}}(t) & \delta_{\mathrm{cmd}}(t) & \tau(t)
    \end{bmatrix}^{\top}
$
with the commanded velocity of the e-scooter at its rear wheel $v_{\mathrm{cmd}}(t)$, the commanded steering angle $\delta_{\mathrm{cmd}}(t)$, and the torque applied to the reaction wheel $\tau(t)$.
Moreover, our application faces the state and input constraints $|\omega(t)|\leq \omega_{\max}$, $|v(t)|\leq v_{\max}$, $|\delta(t)|\leq \delta_{\max}$, and $|\tau(t)|\leq \tau_{\max}$.
Note that the desired velocity $v_{\mathrm{cmd}}(t)$ and steering angle $\delta_\mathrm{cmd}$ are processed by already implemented low-level controllers on the hardware to achieve zero tracking error in $v(t)$ and $\delta(t)$. 
This step is, however, neglected in the following due to its low relevance for the proposed collision avoidance approach.
\vspace*{-0.25\baselineskip}

As discussed in~\cite{Soloperto2021}, the roll dynamics for the balancing of the e-scooter can be decoupled from the remaining dynamics corresponding to the e-scooter's longitudinal and lateral motion.
More precisely, stabilization of the vertical position of the e-scooter is achieved by following the results in~\cite{Wenzelburger2020} based on the inverted pendulum dynamics 
\begin{equation*}
    \begin{bmatrix}
        \dot{\phi}(t)\\
        \ddot{\phi}(t)\\
        \dot{\omega}(t)
    \end{bmatrix} = 
    \begin{bmatrix}
        \dot{\phi}(t)\\
        (m g z_\mathrm{m} \sin(\phi(t)) - \tau(t))/J_\mathrm{e}, \\
        -(m g z_\mathrm{m} \sin(\phi(t)))/J_\mathrm{e} + \tau(t)/J_\mathrm{d}
    \end{bmatrix},
\end{equation*}
where $m$ is the mass of the overall system, $z_\mathrm{m}$ is the height of the center of mass above the ground for $\phi(t)=0$, $g=\SI{9.81}{\meter\per\second^2}$ is the gravity acceleration, and the parameters $J_\mathrm{e}$ and $J_\mathrm{d}$ denote the moments of inertia of the overall system and of the reaction wheel, respectively, both computed with respect to their rotational axes.
Further, the e-scooter's motion can be described by a single-track model, i.e., the kinematic bicycle model~\citep{Kong2015}
\begin{equation*}
    \begin{bmatrix}
        \dot{x}_{\mathrm{p}}(t)\\
        \dot{y}_{\mathrm{p}}(t)\\
        \dot{\psi}(t)
    \end{bmatrix} = 
    \begin{bmatrix}
        v(t) \cos(\psi(t)+\delta(t))/\cos(\delta(t))\\
        v(t) \sin(\psi(t)+\delta(t))/\cos(\delta(t))\\
        v(t) \tan(\delta(t))/l\\
   \end{bmatrix}, 
\end{equation*}
where $l$ denotes the distance from the front to the rear wheel.
Hence, the considered e-scooter can be modeled and controlled via simple differential equations.
\vspace*{-0.4\baselineskip}

\subsection{Hardware setup}\label{sec:hardware-setup}
\vspace*{-0.5\baselineskip}

Here, we briefly describe the hardware setup of the autonomous e-scooter prototype, on which we deploy the proposed collision avoidance safety filter.
The setup is related to the one described in~\cite{Soloperto2021}, but we use a different e-scooter model with additional components for the collision avoidance framework. 
In particular, the prototype is constructed using an \emph{Egret Pro} e-scooter which is augmented by a reaction wheel in order to allow for the stabilization in the roll dynamics (see Fig.~\ref{IMG:HW:EScooter}). 
The stabilization algorithm is implemented on a \emph{VESC 6 MkV} motor controller, which directs a current to a \emph{Maxon EC90 flat} DC motor to actuate the reaction wheel. 
The controller itself is based on the approach proposed in~\cite{Wenzelburger2020} and takes as input an estimation of the roll angle. 
We employ a Mahony filter~\citep{Mahony2008} to compute the roll angle estimate via sensor fusion of the different measurements of the inertial measurement unit of the VESC.
As shown in Fig.~\ref{IMG:HW:EScooter}, steering is achieved by a \emph{PD4-C5918L4204-E-08} stepper motor mounted inside the reaction wheel casing, which in turn is fixed to the deck of the e-scooter. 
An additional \emph{VESC 6 MkV} motor controller is used to actuate the rear wheel for driving forward and backward.
Both VESC motor controllers run at \SI{1}{\kilo\hertz} to enable fast and precise control of the driving velocity and roll angle.
More computationally demanding operations are executed on a \emph{Raspberry Pi 4B} at \SI{50}{\hertz}. 
Tasks on the Raspberry Pi involve, for instance, the detection of obstacles and sending the desired input signals $u(t)$ to the motor controllers. 
The proposed mechanical and electronic design allows us to place most of the added components into a single casing, which encloses the reaction wheel. 
In addition, this configuration is beneficial since it only marginally modifies the original e-scooter, and, therefore, other off-the-shelf e-scooters can be easily upgraded to an autonomous version by adding our hardware and casing.
Further, in order to contain the weight of the e-scooter, we make use of the power supply of its original battery to power the reaction wheel, the steering motor, as well as all the installed electronics. 
In this configuration, the weight of the vehicle increases by~\SI{11.5}{\kilogram}, leading to a total weight of \SI{34}{\kilogram}.

To establish the proposed collision avoidance safety filter of this paper, we require additional components on the e-scooter.
In particular, we employ three low-cost \emph{HC-SR04P} ultrasonic sensors to detect pedestrians and other obstacles in front of the e-scooter. 
As depicted in Fig.~\ref{IMG:HW:ultrasonic}, one ultrasonic sensor is facing straight, and the other two are oriented to the sides of the e-scooter.
On the Raspberry Pi, we execute \emph{Python} code with the \emph{gpiozero} library to measure the distance to the closest object with a frequency of \SI{10}{\hertz}.
We note that the chosen sensor model is a low-cost solution, which is beneficial for the deployment in large fleets of e-scooters or other cost-restricted applications.
This solution differs from other available sensor technologies, such as LiDAR sensors or cameras, which have a greater detection range but are more expensive, require more computing power for post-processing, may be sensitive to lighting conditions, and are more complex to integrate into the e-scooter's hardware.
\vspace*{-0.25\baselineskip}

Each ultrasonic sensor detects objects in a 3D cone with an angle of $15^\circ$, ranging from a minimum distance of~\SI{2}{\centi\meter} to a maximum distance of~\SI{400}{\centi\meter}, with an accuracy of \SI{0.3}{\centi\meter}.
In practice, however, this accuracy is hardly met since the sensors are affected by the environment, e.g., by the material of the detected object, and, thus, measurements may be faulty or missing completely.
Here, the outer ultrasonic sensors are facing both sides of the e-scooter with an angle of $24^\circ$, where each outer sensor is positioned with a distance of \SI{37}{\milli\meter} to the center.
The low requirements on the supplied voltage allows us to connect the ultrasonic sensors directly to the Raspberry Pi, without additional wiring to the power supply of the e-scooter. 
The sensors are fixed to the handlebar by a 3D-printed component which is specifically designed to ensure that the sensor rotates with the handlebar, and, hence, it is always able to detect obstacles in driving direction as well as in the relevant surrounding of the e-scooter, see Fig.~\ref{IMG:HW:ultrasonic}.
Based on the field of view, the height of the frontal sensor is fixed at~\SI{56}{\centi\meter} above the ground while the sensors on the side are fixed at~\SI{50}{\centi\meter} above the ground. 
By this, we ensure that 1) the sensors detect a wide range of obstacles and 2) there is no interference with the ground.
\vspace*{-0.25\baselineskip}
\section{Collision avoidance safety filter}\label{sec:controller}
\vspace*{-0.6\baselineskip}

Next, we detail the proposed collision avoidance safety filter ensuring a safe distance to unknown obstacles.
In particular, we provide an overview of the controller structure (Section~\ref{sec:controller-structure}), describe the employed filter for the distance measurements (Section~\ref{sec:controller-filter}), and detail the design of the distance controller (Section~\ref{sec:controller-design}).
We note that the derived collision avoidance safety filter is independent of the specific dynamics of the e-scooter presented in Section~\ref{sec:system-dynamics} and can be applied to other autonomous vehicles as well.
\vspace*{-0.45\baselineskip}

\subsection{Overview of the controller structure}\label{sec:controller-structure}
\vspace*{-0.6\baselineskip}

To ensure that the e-scooter can autonomously stabilize its upright position, we assume that the balancing algorithm proposed in~\cite{Wenzelburger2020} is active at all times.
Hence, balancing the e-scooter's upright position is not affected by the derived collision avoidance safety filter.
In addition, we assume that a high-level planner is available with the goal to compute the desired velocity $v_{\mathrm{cmd}}(t)$ and the desired steering angle $\delta_{\mathrm{cmd}}(t)$ based on any feasible desired state $x_{\mathrm{des}}(t)$ and the current state $x(t)$.
\begin{figure}[t]
    \centering
    \usetikzlibrary{backgrounds}
\begin{tikzpicture}[scale=1.0,>=latex]
	\node[block, align=center, fill=white,minimum height=1cm,minimum width=1.7cm] (contrDist) {\small Distance\\\small controller};
	\node[block, right=of contrDist,minimum height=2cm,minimum width=1.7cm,yshift=0.5cm] (sys) {\small E-scooter};
	\node[block, left=of contrDist,align=center,minimum height=2cm,minimum width=1.7cm,yshift=0.5cm] (planner) {\small High-level\\\small planner};
	\node[block, below=0.8 of contrDist, fill=white,minimum height=0.5cm,minimum width=1.7cm] (filter) {\small Filter};
	
	\draw[->] (sys) |- (filter) node[pos=0.3,right,align=center]{\small$d_{\mathrm{meas}}^{(\mathrm{c})}(t)$,\\\small$d_{\mathrm{meas}}^{(\mathrm{l})}(t)$,\\\small$d_{\mathrm{meas}}^{(\mathrm{r})}(t)$};
	\draw[->] let \p1 = (contrDist), \p2 = ($(sys.north)+(0,0.25)$) in (sys) |- (\x1,\y2) node[above]{\small$x(t)$} -| (planner);
	\draw[->] ($(planner.south) + (0,-0.95)$) node[below,align=center]{\small$x_{\mathrm{des}}(t)$} -- (planner);
	\draw[->] (filter) -- node[pos=0.5, right] {\small$d_{\mathrm{crit}}(t)$} (contrDist);
	\draw[->] (sys) -- ($(sys.east) + (0.5,0)$) node[pos=0.5,above]{\small$\delta(t)$} |- ($(filter.south) + (0,-0.5)$) -- (filter.south);
	\draw[->] let \p1 = (contrDist), \p2 = (sys.west) in (contrDist) -- node[pos=0.5, above] {\small$v_{\mathrm{safe}}(t)$} (\x2,\y1);
	\draw[->] let \p1 = (planner.east), \p2 = (contrDist) in (\x1,\y2) -- node[pos=0.5, above] {\small$v_{\mathrm{cmd}}(t)$} (contrDist);
	\draw[->] ($(planner.north east)+(0,-0.5)$) -- node[pos=0.5, above] {\small$\delta_{\mathrm{cmd}}(t)$} ($(sys.north west)+(0,-0.5)$);

	\node at ($(filter.south) + (0,-0.05)$) (filTemp) {};
	\node at ($(contrDist.north) + (0,0.05)$) (contrTemp) {};
	\node[left=0.4 of filter,rotate=90,xshift=1.55cm,align=left] {\small Coll. avoidance \\ \small safety filter};
	\node[right=0.3 of filter] (collisionAvoidanceR) {};
	\node[left=0.5 of filter] (collisionAvoidanceL) {};
	\begin{scope}[on background layer]
    	\node[fill=gray!20,fit=(filter) (filTemp) (contrDist) (contrTemp) (collisionAvoidanceL) (collisionAvoidanceR)] {};
    \end{scope}
\end{tikzpicture}
    \vspace*{-0.7\baselineskip}
    \caption{Overview of the employed controller structure for the driving dynamics of the autonomous e-scooter.}
    \label{fig:controller-structure}
\end{figure}
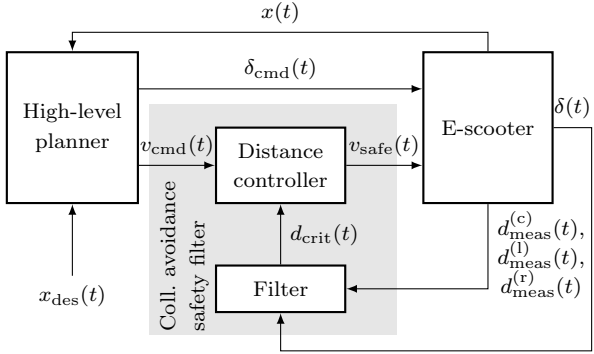
However, the high-level planner cannot guarantee the feasibility of the commanded path as it has no access to local environment information, possibly leading to collisions with unknown obstacles, e.g., pedestrians.
\vspace*{-0.15\baselineskip}

The main goal of this work is to appropriately adjust the desired velocity $v_{\mathrm{cmd}}(t)$ such that collisions are avoided at all times, while simultaneously allowing the e-scooter to deviate as little as possible from the desired behavior planned by the high-level planner.
To this end, we propose a \textit{distance controller} that uses the commanded velocity $v_\mathrm{cmd}(t)$ and a later in~\eqref{eq:critical-distance} described critical distance $d_{\mathrm{crit}}(t)$ between the e-scooter and the closest obstacle to generate a velocity $v_{\mathrm{safe}}(t)\leq v_{\mathrm{cmd}}(t)$ via~\eqref{eq:safe-velocity}.
More precisely, the safe velocity $v_{\mathrm{safe}}(t)$ limits the desired velocity $v_{\mathrm{cmd}}(t)$ under consideration of a detected obstacle to ensure a safe driving behavior of the e-scooter.
Here, we emphasize that the collision avoidance is only active if the commanded velocity is non-negative, i.e., the high-level controller does not command the e-scooter to drive backwards and, thus, away from the obstacle.
An important feature of the proposed approach is the deployed filter which combines the measurements of all three ultrasonic sensors depending on the current steering angle $\delta(t)$ to compute the critical distance $d_{\mathrm{crit}}(t)$.

An overview of the general controller structure related to the driving dynamics is shown in Fig.~\ref{fig:controller-structure}.
The described distance controller together with the measurement filter therefore functions as a safety filter.
It can be easily integrated in other micromobility solutions as Fig.~\ref{fig:controller-structure} demonstrates.
Existing applications only require the incorporation of the proposed safety filter before applying the desired control input to the hardware.
\vspace*{-0.45\baselineskip}

\subsection{Distance measurement filter} \label{sec:controller-filter}
\vspace*{-0.6\baselineskip}

As described in the last section, our designed distance controller receives a critical distance $d_{\mathrm{crit}}(t)$ to the closest obstacle with respect to the e-scooter.
To this end, we use filtered distance measurements $d_{\mathrm{filt}}^{(k)}(t)$ rather than the raw measurements $d_{\mathrm{meas}}^{(k)}(t)$ of the three ultrasonic sensors with the aim to improve the quality and accuracy of the signals and, hence, of the resulting velocity $v_{\mathrm{safe}}(t)$.
Here, $k\in\{\mathrm{c},\ell,\mathrm{r}\}$ corresponds to the center, left, and right ultrasonic sensor, respectively.
Then, the individual filtered distance measurements are combined to compute the critical distance $d_{\mathrm{crit}}(t)$.
\vspace*{-0.25\baselineskip}

\subsubsection{Filter design:}
\begin{figure}[t]
    \centering
    \input{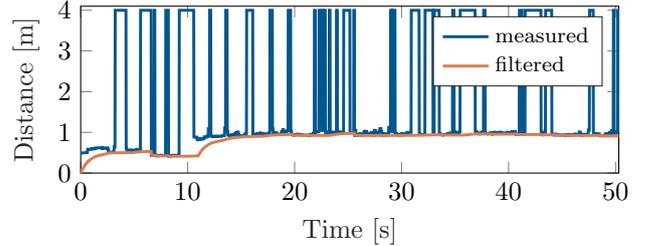}
    \vspace*{-\baselineskip}
    \caption{A sample distance measurement of one ultrasonic sensor and the respective filtered values. As time constants, we choose $T_\mathrm{i} = \SI{0.79}{\second}$ and $T_\mathrm{d} = \SI{0.03}{\second}$.}
    \label{fig:us-filter}
\end{figure}

First, we inspect the sample measurement of one of the ultrasonic filters shown in Fig.~\ref{fig:us-filter} from a real-world experiment, where the e-scooter is driving towards a person.
We observe that the raw measurements are either rather accurate or falsely large due to missed detections, but not falsely small.
Then, we leverage these observations for the filter design of each ultrasonic sensor and implement an exponential smoothing filter
\begin{equation*}
    d_{\mathrm{filt}}^{(k)}(t) = \alpha d_{\mathrm{mem}}^{(k)}(t) + (1-\alpha) d_{\mathrm{filt}}^{(k)}(t-\Delta t)
\end{equation*}
with $k\in\{\mathrm{c},\ell,\mathrm{r}\}$ and factor $\alpha\in (0,1]$ reducing noise in the distance measurements, where
\begin{equation*}
    d_{\mathrm{mem}}^{(k)}(t)
    = \min_{\tau\in[\max\{0,t-\tau_\mathrm{mem}+1\},t]} d_{\mathrm{meas}}^{(k)}(\tau)
\end{equation*}
denotes the minimum of the last $\tau_\mathrm{mem}\geq 1$ measured distance measurements.
More precisely, the introduced memory parameter $\tau_\mathrm{mem}$ mitigates missed detections to ensure that the filtered distance measurements are not erroneously overestimated.
The state of the filter is initialized as $d_{\mathrm{filt}}^{(k)}(0)=0$.
The introduced smoothing filter represents a delayed version of a first-order filter, where the smoothing factor $\alpha=1-e^{-\Delta t/T}$ yields the time constant
$
    T = \nicefrac{-\Delta t}{\ln(1-\alpha)}
$.
Here, $\ln(\cdot)$ denotes the natural logarithm and $\Delta t$ is the sampling frequency of the respective hardware, in our case $\Delta t = \SI{0.02}{\second}$.
The time constant is chosen in dependence of whether the measured distance is increasing or decreasing.
More precisely, we define
\begin{equation*}
    T^{(k)}(t) =
    \begin{cases}
        T_\mathrm{i} & \text{if } d_{\mathrm{meas}}^{(k)}(t) > d_{\mathrm{filt}}^{(k)}(t-\Delta t),    \\
        T_\mathrm{d} & \text{if } d_{\mathrm{meas}}^{(k)}(t) \leq d_{\mathrm{filt}}^{(k)}(t-\Delta t),
    \end{cases}
\end{equation*}
where $T_\mathrm{i}>0$ for increasing measurements is significantly larger than $T_\mathrm{d}>0$ for decreasing measurements.
Note that the memory of size $\tau_\mathrm{mem}$ introduces an additional delay for increasing measurements, which makes the filter more conservative but also more robust to missed detections.
All three constants are tuning variables which depend on the used hardware and the considered velocity range of the vehicle.
Using different time constants for increasing or decreasing distances resembles an important safety feature.
More precisely, if an obstacle is at close range but the sensor receives no or a wrong echo due to a disturbance, then the distance measurement is erroneously large.
Without filtering, no obstacle would be detected for a short period of time which would result in the e-scooter accelerating forward until the next correct measurement is taken.
However, by using the above described filter parameters, we are still able to employ a safety-conscious solution.
On the one hand, the proposed approach requires multiple measurements of an obstacle-free path to recover from halting due to an obstacle.
On the other hand, only a few measurements indicating an obstructed path are required to indeed detect an obstacle.

For the sample measurement in Fig.~\ref{fig:us-filter}, where in many but short time intervals the measured distance $d_{\mathrm{meas}}(t)$ is falsely large due to missed detections, the filtered distance $d_{\mathrm{filt}}(t)$ ensures safe operation on the cost of a more conservative behavior with respect to the detected obstacle.
More precisely, the filtered signal follows decreasing tendencies fast while it delays increasing distance measurements to account for possibly missed detections.
We emphasize that the proposed filter results in an accurate and smooth distance measurement, where the resulting delay is outweighed by the obtained safety benefits of the filter.
\vspace*{-0.25\baselineskip}

\subsubsection{Critical distance:}
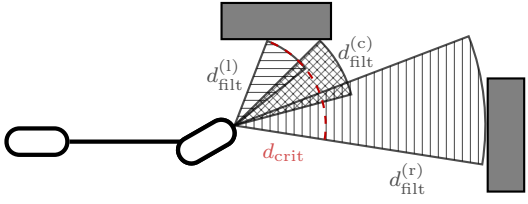
\begin{figure}[t]
    \centering
    \tikzset{
pill/.style={minimum width=0.8cm,minimum height=3.5mm,rounded corners=1.7mm,draw}
}
\begin{tikzpicture}[scale=1.2]
	\colorlet{darkRed}{red!75!black}

	\coordinate (M) at (-3,0);
	\draw[thick, black, pattern=horizontal lines, pattern color=black,opacity=0.7] ([shift=(69:1cm)]M) arc (69:39:1cm) node[pos=0](left){} -- (M) -- cycle node[pos=0.6,left]{\small$d_\mathrm{filt}^{(\mathrm{l})}$};
	\draw[thick, black, pattern=vertical lines, pattern color=black,opacity=0.7] ([shift=(-9:2.75cm)]M) arc (-9:21:2.75cm) node[pos=0](right){} -- (M) -- cycle node[pos=0.7,below]{\small$d_\mathrm{filt}^{(\mathrm{r})}$} node[pos=0.2,below,darkRed]{\small$d_\mathrm{crit}$};
	\draw[thick, black, pattern=crosshatch, pattern color=black,opacity=0.7] ([shift=(15:1.325cm)]M) arc (15:45:1.325cm) node[pos=0.5,above right,xshift=-0.15cm,yshift=-0.15cm]{\small$d_\mathrm{filt}^{(\mathrm{c})}$} node[pos=0.7](center){} -- (M) -- cycle;

	\node[anchor=east,ultra thick,pill,rotate=30] (front) at (M.west){};
	\draw[ultra thick] ($(front) + (-1.5,0)$) node[left,pill]{} -- (front.center);
	\node[ultra thick,pill,rotate=30,fill=white] at (front){};

	\draw[black,fill=gray,thick] ($(right)+(0.06,-0.3)$) rectangle ++(0.4,1.25);
	\draw[black,fill=gray,thick] ($(left)+(-0.5,0.02)$) rectangle ++(1.2,0.4); 

	\draw[darkRed,thick,dashed] ([shift=(-9:1cm)]M) arc (-9:69:1cm);

\end{tikzpicture}
    \vspace*{-1\baselineskip}
    \caption{Filtered measurement cones of each ultrasonic sensor and the resulting critical distance.}
    \label{fig:distances-circular}
\end{figure}

Recall that we use three ultrasonic sensors to enlarge the field of view.
In order to compress the information of the three sensors into a single critical distance $d_{\mathrm{crit}}(t)$, we define
\begin{equation}\label{eq:critical-distance}
    d_{\mathrm{crit}}(t) = \min\{d_{\mathrm{filt}}^{(\mathrm{c})}(t), d_{\mathrm{filt}}^{(\ell)}(t), d_{\mathrm{filt}}^{(\mathrm{r})}(t)\}.
\end{equation}
Then, the critical distance ensures safe operation of the e-scooter regardless of the desired steering angle, since it denotes always the smallest distance to an obstacle in the e-scooter's field of view, see Fig.~\ref{fig:distances-circular}.

\begin{rem}\label{rem:critical-distance-weighting}
    The minimization in~\eqref{eq:critical-distance} is a simple approach to compress the information of the three ultrasonic sensors to a single value for the critical distance.
    We emphasize that more sophisticated approaches, e.g., using a weighted average of the three distances, are possible and can lead to a more accurate representation of the actual environment.
    In particular, the combined implementation of a weighted average and the minimum of the distances of all three sensors offers an effective way to include the steering angle in the computation of the critical distance, and, hence, allows the e-scooter to consider only relevant obstacles by ignoring obstacles which are not important for its imminent path.
    However, we note that the used weighting needs to be carefully chosen to ensure that the e-scooter is able to detect and react to unknown obstacles in time.
\vspace*{-0.25\baselineskip}
\end{rem}

\subsection{Controller design} \label{sec:controller-design}
\vspace*{-0.4\baselineskip}

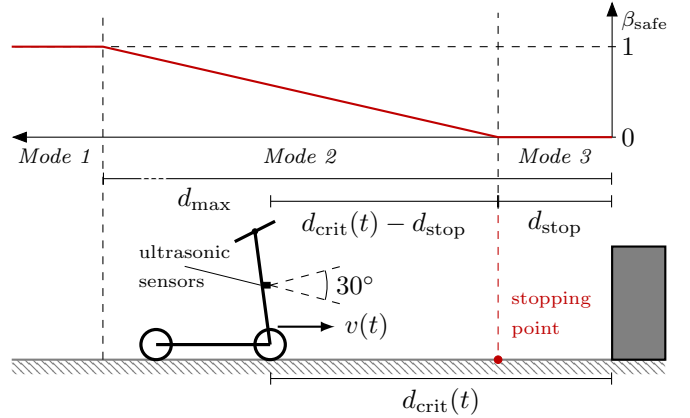
\begin{figure}[t]
    \centering\hspace*{-0.25cm}
    \usetikzlibrary{quotes,angles}
\begin{tikzpicture}[scale=1]
	
	\tikzstyle{ground}=[fill,pattern=north east lines,draw=none,minimum width=1.0,minimum height=0.1]
	\colorlet{darkRed}{red!75!black}
	
	\node[] (scooter) at (2,0) {};
	\node[] (obstacle) at (6.5,0) {};

	\draw[white,pattern=north west lines, pattern color=gray] ($(scooter.center)-(3.4,0)$) rectangle ($(obstacle.center)+(0.7,-0.2)$);
	\draw[gray,thick] ($(scooter.center)-(3.4,0)$) -- ($(obstacle.center)+(0.7,0)$);
	\node[label={[align=left,xshift=0.7cm,yshift=0.05cm]\scriptsize{\color{darkRed}{stopping}}\\\scriptsize{\color{darkRed}{point}}}] (stopPoint) at ($(obstacle)+(-1.5,0)$) {};
	\draw[darkRed,fill=darkRed] (stopPoint) circle [radius=0.05];
	\draw[darkRed,dashed] (stopPoint) -- ++(0,2);

	\def\wheelRadius{0.4}
	\begin{scope}[scale=0.5]
		\begin{scope}[shift={(0,\wheelRadius)}]
			\node[] (leftWheel) at (1,0) {};
			\node[] (rightWheel) at ($(leftWheel)+(3,0)$) {};
			\draw[black,very thick] (leftWheel) circle [radius=\wheelRadius];
			\draw[black,very thick] (rightWheel) circle [radius=\wheelRadius];
			\draw[black,very thick] (leftWheel.center) -- (rightWheel.center);
			
			\node[] (handleBarTop) at ($(leftWheel)+(2.6,3)$) {};
			\draw[black,very thick] (handleBarTop.center) -- (rightWheel.center);
			\draw[black,very thick] (handleBarTop.center) -- ++(0.53,0.27);
			\draw[black,very thick] (handleBarTop.center) -- ++(-0.53,-0.27);
			\draw[black,fill=black] (handleBarTop) circle [radius=0.07];
			
			\def\xSizeUsSensor{0.20}
			\def\ySizeUsSensor{0.15}
			\node[] (usMounting) at ($(rightWheel)!0.5!(handleBarTop)$) {};
			\node[] (usSensor) at ($(usMounting)+(\xSizeUsSensor,\ySizeUsSensor/2)$) {};
			\draw[black,fill=black] (usMounting) rectangle ++(\xSizeUsSensor,\ySizeUsSensor);
			
			\draw[black,dashed] (usSensor.center) -- +(15:2cm) coordinate (a);
			\draw[black,dashed] (usSensor.center) -- +(-15:2cm) coordinate (b);
			\draw pic["\ang{30}",draw,solid,angle radius=.75cm,angle eccentricity=1.5] {angle = b--usSensor--a};
			
			\draw[black, very thin] (usSensor.west) -- node[pos=1.0, align=left] {\scriptsize{ultrasonic}\\\scriptsize{sensors}} +(-2,0.5);
			
			\draw[-latex,thick] ($(usSensor.center)+(0.2,-1.1)$) -- ++(1.5,0) node[right] {$v(t)$};
		\end{scope}
	\end{scope}

	\begin{scope}[shift={(obstacle)},scale=1.0]
		\draw[black,fill=gray,thick] (0,0) rectangle ++(0.7,1.5);
	\end{scope}

	\def\yPositioningLongDistance{-0.25}
	\def\yPositioningShortDistances{1.05}
	\def\yPositioningMaxDistance{1.2}
	\draw[|-|] ($(scooter.center)+(0,\yPositioningLongDistance)$) -- ($(obstacle.center)+(0,\yPositioningLongDistance)$) node[midway,below] {$d_{\mathrm{crit}}(t)$};
	\draw[|-|] ($(scooter.center)+(0,2*\yPositioningShortDistances)$) -- ($(stopPoint.center)+(0,2*\yPositioningShortDistances)$)node[midway,below] {$d_\mathrm{crit}(t)-d_\mathrm{stop}$};
	\draw[|-|] ($(stopPoint.center)+(0,2*\yPositioningShortDistances)$) -- ($(obstacle.center)+(0,2*\yPositioningShortDistances)$)node[midway,below] {$d_\mathrm{stop}$};
	\draw[|-|] ($(scooter.center)+(-2.2,2*\yPositioningMaxDistance)$) -- ($(obstacle.center)+(0,2*\yPositioningMaxDistance)$) node[pos=0.2,below] {$d_\mathrm{max}$} node[pos=0.1,color=white]{....};

	\draw[dashed] let \p1 = (scooter.center), \p2 = (stopPoint) in ($(stopPoint)+(0,2)$) -- ($(\x2,\y1)+(0,2.95) + (0,1.65)$);
	\draw[dashed] ($(scooter.center)+(-2.2,0)$) -- ($(scooter.center)+(-2.2,2.95) + (0,1.65)$);
	\path (5.75,2.7) node[above,anchor=center]{\small\emph{Mode 3}};
	\path (2.4,2.7) node[above,anchor=center]{\small\emph{Mode 2}}; 
	\path (-0.2,2.7) node[above left,anchor=east]{\small\emph{Mode 1}};
	\draw[-{Triangle[length=2mm]}] let \p1 = (scooter.center), \p2 = (obstacle.center) in ($(\x2,\y1)+(0,2.95)$) -- ($(0,\y1)+(-1.4,2.95)$);
	\draw[-{Triangle[length=2mm]}] let \p1 = (scooter.center), \p2 = (obstacle.center) in ($(\x2,\y1)+(0,2.95)$) -- ($(\x2,\y1)+(0,2.95)+(0,1.8)$) node[pos=1,below right]{\small$\beta_\mathrm{safe}$} node[pos=0,right]{$0$};
	\path let \p1 = (scooter.center), \p2 = (obstacle.center) in ($(\x2,\y1)+(0,2.95)+(0,1.2)$) node[right]{1};
	\draw[dashed] let \p1 = (scooter.center), \p2 = (obstacle.center) in ($(\x2,\y1)+(0,2.95)+(0.05,1.2)$) -- ($(0,\y1)+(-1.4,2.95) + (0,1.2)$);
	\draw[-] let \p1 = (scooter.center), \p2 = (obstacle.center) in ($(\x2,\y1)+(0,2.95)$) -- ($(\x2,\y1)+(0.05,2.95)$);
	\draw[color=darkRed,thick] let \p1 = (scooter.center), \p2 = (obstacle.center) in ($(\x2,\y1)+(0,2.95)$) -- ($(\x2,\y1)+(-1.5,2.95)$) -- ($(scooter.center)+(-2.2,2.95)+(0,1.2)$) -- ($(0,\y1)+(0,2.95)+(-1.4,1.2)$);
	
\end{tikzpicture}
    \vspace*{-2.2\baselineskip}
    \caption{Overview of the relevant distance variables and the different controller modes.}
    \label{fig:distances}
\end{figure}

For the desired collision avoidance, the maximum distance in which the e-scooter detects obstacles and, hence, reacts to them is denoted by $d_{\mathrm{max}}>0$.
This threshold defines when the ultrasonic sensors are no longer effective or when the e-scooter no longer needs to react to distant obstacles.
Further, we require the e-scooter to decelerate and potentially stop at a predefined safe stopping distance $d_\mathrm{stop}$ to obstacles, where $0\leq d_\mathrm{stop}< d_\mathrm{max}$.
The difference between the obtained critical distance $d_{\mathrm{crit}}(t)$ and the stopping distance $d_\mathrm{stop}$ describes the remaining distance that is left for the e-scooter before reaching the stopping point, see Fig.~\ref{fig:distances}.
The goal of the proposed controller is to ensure collision avoidance at all times.
Hence, we need to ensure that $d_{\mathrm{crit}}(t) > 0$ for all $t\geq 0$, and, if possible, even $d_{\mathrm{crit}}(t)\geq d_\mathrm{stop}$ such that we always have a positive margin to the obstacle or even to the stopping point.
In order to fulfill these objectives, we distinguish three different situations depending on the critical distance $d_{\mathrm{crit}}(t)$.
Recall that the collision avoidance is only active if the high-level controller demands the e-scooter to drive forward.

\begin{procedure}\label{proc:safe-velocity}
    We define the safe velocity as 
    \begin{equation}\label{eq:safe-velocity}
        v_\mathrm{safe}(t) = \min\{\beta_\mathrm{safe}(d_\mathrm{crit}(t)) v_\mathrm{cmd}(t),v_\mathrm{cmd}(t)\}
    \end{equation}
    with
    \vspace*{-0.4\baselineskip}
    \begin{equation*}
        \beta_\mathrm{safe}(d_\mathrm{crit}(t)) =
        \begin{cases}
            1                                                                           & \text{if } d_\mathrm{crit}(t) > d_\mathrm{max},                           \\
            \frac{d_{\mathrm{crit}}(t)-d_\mathrm{stop}}{d_\mathrm{max}-d_\mathrm{stop}} & \text{if } d_\mathrm{stop} \leq d_{\mathrm{crit}}(t) \leq d_\mathrm{max}, \\
            0                                                                           & \text{if } d_{\mathrm{crit}}(t) < d_\mathrm{stop},                        \\
        \end{cases}
    \end{equation*}
    where $\beta_\mathrm{safe}(d_\mathrm{crit}(t))\in[0,1]$ and the minimization in~\eqref{eq:safe-velocity} ensures that active braking of the high-level controller via negative velocities $v_\mathrm{cmd}(t) < 0$ is not affected by the collision avoidance.
    Then, applying the safe velocity ensures that the e-scooter is able to safely stop in front of obstacles such that collisions are avoided at all times.
    \vspace*{-0.25\baselineskip}
\end{procedure}
The different modes of the collision avoidance scheme via the defined $\beta_\mathrm{safe}(d_\mathrm{crit}(t))$ in Procedure~\ref{proc:safe-velocity} are illustrated in Fig.~\ref{fig:distances} and detailed in the following, where we focus on the case of nonnegative commanded velocities.

\emph{Mode 1: $d_\mathrm{crit}(t) > d_\mathrm{max}$.~}
We start with the simplest mode, where the e-scooter is far away from any obstacles.
Hence, there is no need to limit the commanded velocity.
Thus, $\beta_\mathrm{safe}(d_\mathrm{crit}(t))=1$ leads to $v_{\mathrm{safe}}(t) = v_{\mathrm{cmd}}(t)$ ensuring that the controller does not interfere with the velocity chosen by the high-level controller.
Consequently, the collision avoidance safety filter is virtually inactive.
\vspace*{-0.25\baselineskip}

\emph{Mode 2: $d_\mathrm{stop} \leq d_{\mathrm{crit}}(t) \leq d_\mathrm{max}$.~}
In this case, the e-scooter is still far enough away from the detected obstacle, i.e., at least further away than the defined stopping distance $d_\mathrm{stop}$.
However, the sensors detect an obstacle and, therefore, a collision is imminent if the velocity is not adjusted accordingly.
Thus, the objective within this distance interval is to ensure that the e-scooter decelerates such that safety can be ensured.
This behavior is established through the factor $\beta_\mathrm{safe}(d_\mathrm{crit}(t))=\frac{d_{\mathrm{crit}}(t)-d_\mathrm{stop}}{d_\mathrm{max}-d_\mathrm{stop}}\in[0,1]$, i.e., the applied safe velocity is a simple multiplication of the commanded velocity with a linear scaling depending on the remaining distance to the stopping point.
\vspace*{-0.25\baselineskip}

\emph{Mode 3: $d_{\mathrm{crit}}(t) < d_\mathrm{stop}$.~}
If the e-scooter is closer to the obstacle than the defined stopping distance $d_\mathrm{stop}$, it must stop abruptly to ensure safety.
To this end, the safe velocity $v_{\mathrm{safe}}(t)$ is set to zero by $\beta_\mathrm{safe}(d_\mathrm{crit}(t))=0$, and, hence, the e-scooter is commanded to stop immediately.
\vspace*{-0.25\baselineskip}

We emphasize that the proposed controller structure is a simple and intuitive way to ensure collision avoidance for autonomous vehicles relying only on low-cost ultrasonic sensors and noisy distance measurements.
In particular, this design guide should also allow practitioners without expert knowledge in control theory to apply the approach. 
Therefore, the provided method is suitable for a wide range of applications and users.
The resulting behavior of the collision avoidance safety filter can be further improved by using more sophisticated control strategies, e.g., a proportional-integral-derivative controller ensuring that the e-scooter is able to retain a fixed distance to the obstacle, which is left for future work.
However, already the introduced simple scheme computes a safe velocity $v_{\mathrm{safe}}(t)$ to ensure that the e-scooter safely stops in front of obstacles such that collisions are avoided at all times (see Section~\ref{sec:experiments} for real-world experiments).
Here, the defined stopping distance $d_\mathrm{stop}$ is crucial for the resulting safety.
It needs to be chosen carefully based on the hardware, the application, and the maximum velocity to ensure that the e-scooter can detect and react to obstacles in time.

\begin{rem}
    The proposed controller structure is designed to ensure that the e-scooter is able to safely stop in front of obstacles such that collisions are avoided at all times.
    More precisely, the collision avoidance safely stops the e-scooter to wait until the obstacle potentially moves away and is, thus, no longer detected.
    During this time, the e-scooter may only drive backwards if commanded by the high-level controller, but not if the obstacle is closer to the e-scooter than $d_\mathrm{stop}$.
    This is crucial since the e-scooter cannot detect obstacles behind it, making collision within its blind spots possible.
    However, our setting can be easily extended to enable safe reverse driving by installing additional perception sensors facing the backwards direction.
    \vspace*{-0.5\baselineskip}
\end{rem}
\section{Experiments} \label{sec:experiments}
\vspace*{-0.7\baselineskip}

In this section, we demonstrate the effectiveness of the proposed collision avoidance safety filter by presenting three real-world experiments.
For the first experiment (Section~\ref{sec:experiments-straight}), we demonstrate the collision avoidance scheme when driving straight with a static obstacle.
The second experiment illustrates the proposed framework when approaching obstacles in curves (Section~\ref{sec:experiments-curves-min}). 
The behavior of the e-scooter with the proposed safety filter for moving obstacles is shown in the third experiment (Section~\ref{sec:experiments-straight-moving}).
\vspace*{-0.25\baselineskip}

In all experiments, the e-scooter drives autonomously towards either a static or a moving obstacle, while self-balancing its upper equilibrium.
For the implementation of the proposed collision avoidance safety filter, we choose the stopping distance $d_\mathrm{stop} = \SI{0.5}{\meter}$ and the maximum distance $d_\mathrm{max} = \SI{2}{\meter}$ below which obstacle measurements are taken into account.
For filtering the distance measurements $d_{\mathrm{meas}}^{(k)}(t)$, $k\in\{\mathrm{c},\ell,\mathrm{r}\}$, of the three ultrasonic sensors, we use the time constants $T_\mathrm{i} = \SI{0.79}{\second}$ and $T_\mathrm{d} = \SI{0.03}{\second}$ for increasing and decreasing measurements, respectively.
Further, all following plotted velocities are filtered with a moving average filter to remove measurement noise for better interpretability, where we use a window size of $20$ samples centered around the current time step.\footnote{A video recording of all experiments is available at \url{https://www.ist.uni-stuttgart.de/research/Files/CollisionAvoidance.mp4}, where we acknowledge the support by the SC SimTech.}
\vspace*{-0.45\baselineskip}

\subsection{Driving straight}\label{sec:experiments-straight}
\vspace*{-0.6\baselineskip}

In the first experiment, we showcase the detection of static obstacles when driving straight. 
To this end, the performed experiment is set up as follows. 
The high-level planner commands the e-scooter to drive forward with $v_{\mathrm{cmd}}(t) = \SI{1}{\meter\per\second}$ and $\delta_{\mathrm{cmd}}(t) = 0$.
Then, the ultrasonic sensors detect an obstacle such that the e-scooter needs to decelerate and stop in front of the obstacle.
To this end, we apply the safe velocity as described in Section~\ref{sec:controller}.

\begin{figure}
	\centering
	\input{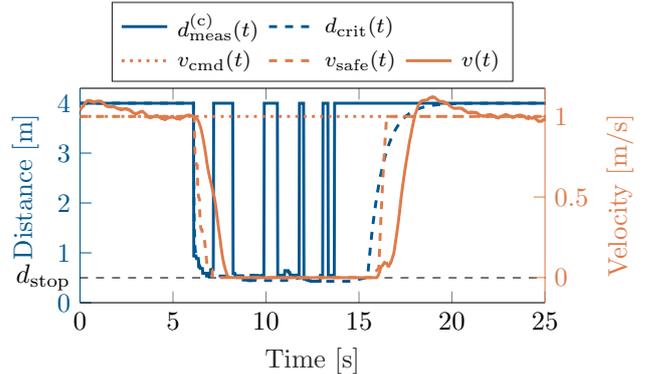} 
    \vspace*{-1.25\baselineskip}
	\caption{Measurements for the experiment in Section~\ref{sec:experiments-straight}.}
	\label{fig:experiment-straight}
	\vspace*{-0.45\baselineskip}
\end{figure}

The resulting measured distance and velocity of the e-scooter together with the computed critical distance, the commanded velocity and the safe velocity are shown in Fig.~\ref{fig:experiment-straight}.
After $\SI{6}{\second}$ of driving straight, the ultrasonic sensors detect an obstacle that is within $d_\mathrm{max} = \SI{2}{\meter}$ and, thus, the collision avoidance is activated and limits the safe velocity accordingly.
The designed controller is able to reduce the velocity of the e-scooter such that it stops safely at the predefined stopping distance $d_\mathrm{stop}$.
Without our employed collision avoidance safety filter, the e-scooter would collide with the obstacle due to the high level planner not accounting for the obstacle and commanding a forward velocity $v_{\mathrm{cmd}}(t) > 0$.
After the obstacle disappears from the e-scooter's path, the e-scooter accelerates again to its commanded velocity $v_{\mathrm{cmd}}(t)$ and follows the commands of the high-level planner.
We emphasize again that the exponentially smoothing filter is crucial for the successful implementation of the collision avoidance safety filter, as it mitigates the faulty distance measurements of the ultrasonic sensors.
As seen in the plot, this comes at the cost of a small delay before acceleration.
\vspace*{-0.5\baselineskip}

\subsection{Driving curves}\label{sec:experiments-curves-min}
\vspace*{-0.65\baselineskip}
Our second experiment incorporates a curved path of the e-scooter.
More precisely, the high-level planner commands the e-scooter to drive forward with $v_{\mathrm{cmd}}(t) = \SI{0.8}{\meter\per\second}$ and $\delta_{\mathrm{cmd}}(t) = \SI{0.4}{\radian}$.
At the same time, the e-scooter needs to avoid any collision with occurring obstacles.

\begin{figure}
	\centering
	\input{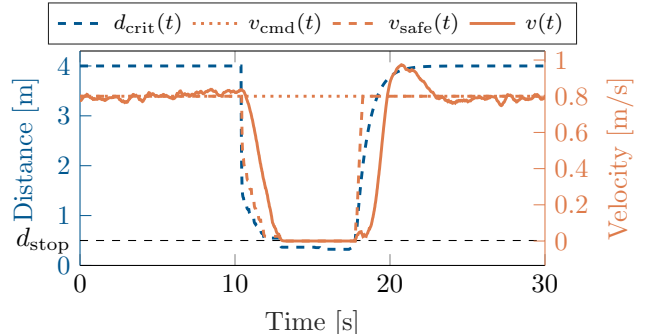} 
    \vspace*{-1.25\baselineskip}
	\caption{Measurements for the experiment in Section~\ref{sec:experiments-curves-min}.}
	\label{fig:experiment-curve-min}
    \vspace*{-0.1\baselineskip}
\end{figure}

Measurements of the resulting e-scooter behavior are depicted in Fig.~\ref{fig:experiment-curve-min}. 
Here, we omit the measured distances to the obstacle of the individual sensors for better readability, and only show the computed critical distance.
For the first \SI{10}{\second} of the experiment, the e-scooter drives in a circle with $v_\mathrm{cmd}(t)$ and $\delta_\mathrm{cmd}(t)$.
Then, an obstacle is detected in the range of one of the three ultrasonic sensors, such that the controller adapts the velocity of the e-scooter via the computed safe velocity according to Section~\ref{sec:controller-design}.
Although the measured distance to the obstacle is partially faulty, the e-scooter is able to safely stop in front of the obstacle and continues its commanded path after the obstacle moves away.
Notably, the e-scooter moves only forward after the obstacle passed all ultrasonic sensors as the critical distance is computed as the minimum of the three filtered distance measurements.
\vspace*{-0.45\baselineskip}

\subsection{Driving straight with moving obstacles}\label{sec:experiments-straight-moving}
\begin{figure}
	\centering
	\input{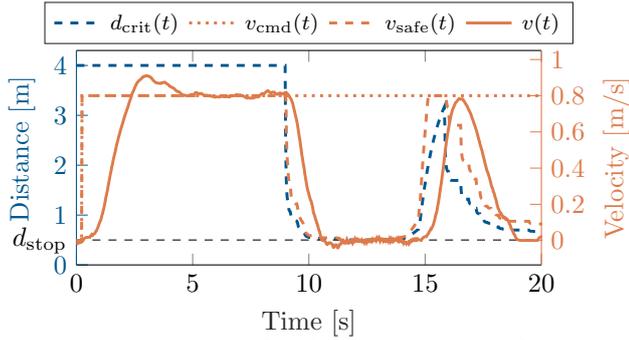} 
    \vspace*{-1.2\baselineskip}
	\caption{Measurements for the experiment in Section~\ref{sec:experiments-straight-moving}.}
	\label{fig:experiment-straight-moving}
\end{figure}

\vspace*{-0.6\baselineskip}
For the third experiment, we consider the case of moving obstacles. 
The setup is related to the first experiment, but the obstacle is now moving and crosses the e-scooter's path unexpectedly.
More precisely, the e-scooter drives forward with $v_{\mathrm{cmd}}(t) = \SI{0.8}{\meter\per\second}$ and $\delta_{\mathrm{cmd}}(t) = 0$.
The results of the experiment are shown in Fig.~\ref{fig:experiment-straight-moving}.
Here, the overall behavior is similar to the experiment in Section~\ref{sec:experiments-straight}, but the e-scooter has to react to moving obstacles which may be detected only shortly before collision. 
As illustrated by the measurements, the e-scooter safely stops in front of the obstacle and accelerates after the obstacle has passed the e-scooter's path. 
After acceleration, the deployed collision avoidance detects another obstacle and safely stops the e-scooter again.
\vspace*{-0.25\baselineskip}

Thus, the collision avoidance safety filter allows the e-scooter to safely navigate in an environment with unknown obstacles, where the proposed approach can be tuned by adjusting the computation of the critical distance according to the specific application.
\vspace*{-0.25\baselineskip}
\section{Conclusion}\label{sec:conclusion}
\vspace*{-0.6\baselineskip}

In this work, we discussed a collision avoidance safety filter for an autonomous e-scooter based on noisy and occasionally missing distance measurements of low-cost ultrasonic sensors.
This procedure is particularly useful as an additional safety-component complementing an existing driving velocity controller.
The proposed safety filter proved useful to stop in front of obstacles, where the formulated collision avoidance is able to ignore obstacles not blocking the e-scooter's imminent path. 
Moreover, the provided hardware and software details in this design guide allow for an easy transfer of the cost-efficient solution relying on low-cost ultrasonic sensors to other autonomous vehicles in the micromobility sector.
The presented approach is designed such that the tuning is intuitive without requiring an extensive background in control theory.
The shown real-world experiments demonstrated the effectiveness of our approach when driving straight as well as driving curves.


\vspace*{-0.5\baselineskip}


\bibliography{e-scooter}             

\end{document}